\pdfoutput=1
\RequirePackage{ifpdf}
\documentclass{JINST}

\usepackage{amsmath}

\title{Compton Backscattering for the Calibration of KEDR Tagging System}
\author{V.V.~Kaminskiy$^{a,b}$\thanks{Corresponding author.}, %
N.Yu.~Muchnoi$^{a,c}$, %
and V.N.~Zhilich$^{a,c}$\\
\llap{$^a$}Budker Institute of Nuclear Physics, 630090 Novosibirsk, Russia\\
\llap{$^b$}Novosibirsk State Technical University, 630073 Novosibirsk, Russia\\
\llap{$^c$}Novosibirsk State University, 630090 Novosibirsk, Russia\\
E-mail: \email{V.V.Kaminskiy@inp.nsk.su}}

\abstract{
KEDR detector has the tagging system (TS) to study the gamma-gamma processes.
To determine the two-photon invariant mass, the energies of the scattered at small angles electrons and positrons are measured by the magnetic spectrometer embedded into the lattice of the VEPP--4M collider.
The energy resolution (scattered electron/positron energy resolution divided by the beam energy) of this spectrometer varies from 0.6\% to 0.03\% depending on the electron/positron energy.
The Compton backscattering of laser radiation on the electron/positron beam is used for the accurate energy scale and resolution calibration of the tagging system.
The report covers the design, recent results and current status of the KEDR TS calibration system.
}

\keywords{Detector alignment and calibration methods}

\begin{document}

\section{KEDR Tagging System}

Two-photon processes $e^-e^+ \rightarrow e^-e^+\gamma^*\gamma^* \rightarrow e^-e^+X $ study is an essential part of the KEDR detector physical program \cite[pp.~1--2]{KEDR}.
For this purpose KEDR has a unique tagging system (TS) for electrons and positrons scattered at small angles ("double-tag" approach) \cite[pp.~29--37]{KEDR}. The invariant two-photon mass can be calculated from the energies $E_{e^-}$, $E_{e^+}$ of the scattered electron and positron:
\begin{equation}
W_{\gamma^*\gamma^*}^2\approx 4(E_{beam}-E_{e^-})(E_{beam}-E_{e^+}).
\label{2g_mass}
\end{equation}
TS consists of the symmetrical magnetic focusing spectrometer (four dipole and four quadrupole magnets, also correctors, solenoids and weak nonlinear elements) embedded into the VEPP--4M collider lattice (see Fig.~{\ref{rsse_layout}}). Scattered electrons/positrons are registered by 8 coordinate detectors, each of them consists of 2-dimensional GEM detector and drift tubes hodoscope. 
Energy range of the spectrometer is 0.39...0.98 of the beam energy; in details, the subsystem TS1 covers the energy range 0.39...0.59, TS2 covers 0.60...0.72, TS3 detects 0.73...0.85, and TS4 detects 0.87...0.98. The energy resolution for scattered electrons (or positrons, briefly, SE) is estimated as 0.6\%...0.03\% of the beam energy. Also the facility is equipped with two BGO crystal calorimeters, intended for detection of photons emitted at small angles.
The resolution for the invariant two-photon mass is better than 1.3\% and it is mostly defined by the beam angular spread.

To gain the proposed performance of TS, a precise energy calibration is needed. 
An energy calibration of TS can be done using Compton backscattering of laser radiation on the beam. Direct calibration of some TS coordinate detectors can be done with Compton electrons/positrons. The rest of TS coordinate detectors can be calibrated using energy-coupled bremsstrahlung electrons/positrons and photons, and BGO calorimeters, calibrated with Compton photons.

\section{Concept}

When a low-energy photon with energy $\omega_0$ collides with an ultrarelativistic electron (or positron) with energy $\varepsilon$, the scattered photon and electron (or positron) have energies strictly dependent on the scattering angle, and particles go within a narrow cone along the initial electron momentum. This particular case is named ``Compton backscattering''.

For head-on collision the minimum energy $\varepsilon_{min}$ of scattered electron (positron) and the maximum energy $\omega_{max}$ of scattered photon are given by:
\begin{eqnarray}
\varepsilon_{min} = \frac{\varepsilon}{1+\lambda} & & \text{and} \label{emin} \\
\omega_{max} = \frac{\varepsilon \lambda}{1+\lambda} \approx 4 \gamma^2 \omega_0~, & & \text{where} \label{wmax} \\
\lambda = \frac{4 \omega_0 \varepsilon}{m^2}~.
\end{eqnarray}

Here $c=1$, $m$ is electron rest energy and $\gamma = \varepsilon/m$, electron Lorentz-factor.

\begin{figure}[htb]
\centering
\includegraphics[width=1\textwidth]{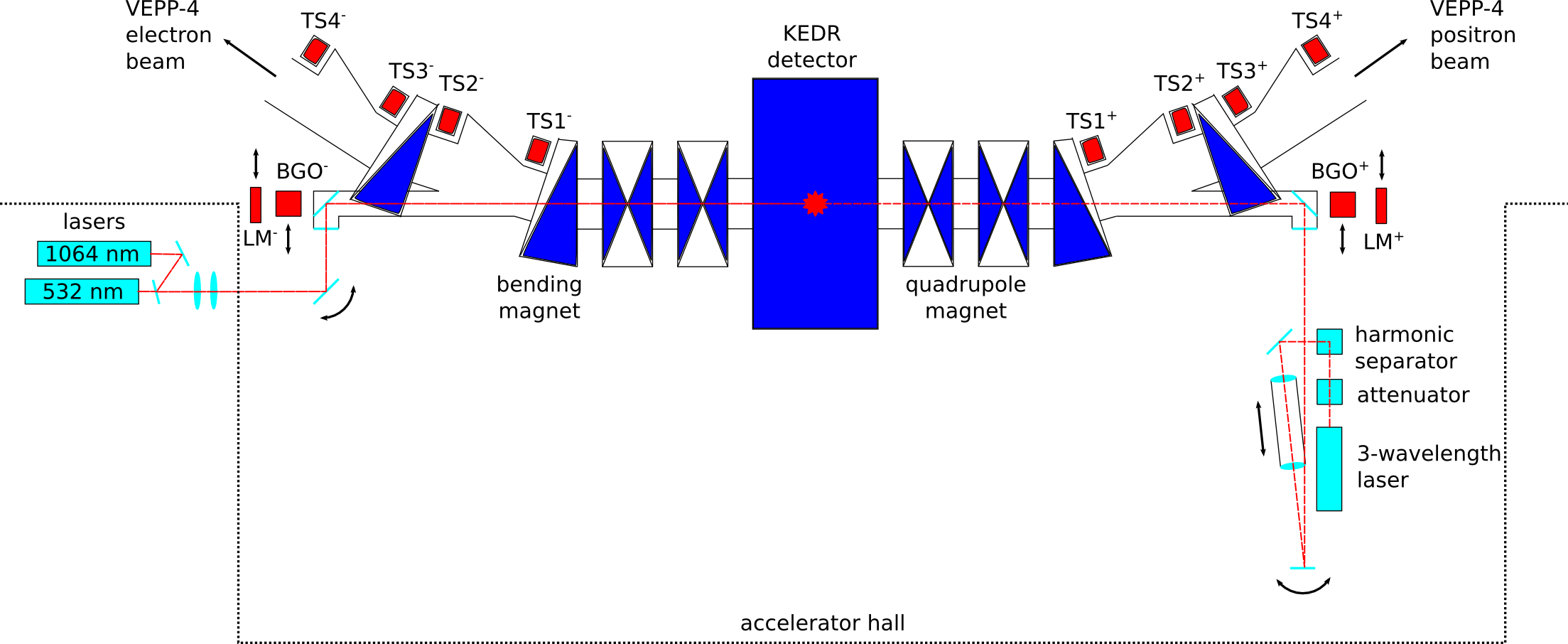}
\caption{Layout of the experiment}
\label{rsse_layout}
\end{figure}

The main idea of TS calibration by Compton backscattering is the following. The laser radiation of few wavelengths is inserted into the vacuum chamber of the VEPP--4M collider to interact with the electron and/or positron beam. The laser radiation is pulsed, the length of the pulse is less than 4~m. Compton interaction occurs within 2~m at the centre of the KEDR detector. Compton scattered electrons and positrons are detected by the coordinate detectors of the TS and Compton backscattered photons are detected by the BGO calorimeters. So, the TS subsystems are calibrated using the minimum SE energy, \ref{emin}, and the calorimeter is calibrated using the maximum photon energy, \ref{wmax}. After the calibration of the BGO calorimeters, TS can be calibrated at any energy using energy-tagged beam-beam or beam-gas bremsstrahlung electrons/positrons or Compton electrons/positrons. Also the energy resolution of TS and the calorimeters can be measured using Compton backscattering. The Figure~\ref{rsse_layout} shows the layout of the installation. 

\section{Details of the Experiment}

\subsection{Lasers}

At the electron side of TS two Nd:YAG Q-switched lasers are installed: 1064~nm (or 1.17~eV photon energy, fundamental harmonic, DTL-324QT by Laser-export Co. Ltd.) of 100~$\mu$J pulse energy \cite{DTL-324QT} and 532~nm (or 2.33~eV, second harmonic, DTL-314QT by Laser-export Co. Ltd.) of 25~$\mu$J pulse energy \cite{DTL-314QT} (at 1~kHz repetition rate). 

At the positron side of TS one 3-wavelength Nd:YLF Q-switched laser (DTL-394QT by Laser-export Co. Ltd.) \cite{DTL-394QT} is installed. It emits 1053~nm radiation (or 1.18~eV, fundamental harmonic) of 100~$\mu$J pulse energy, 527~nm radiation (or 2.35~eV, second harmonic) of 100~$\mu$J pulse energy, and 263~nm radiation (or 4.72~eV, fourth harmonic) of 10~$\mu$J pulse energy (at 1~kHz repetition rate). All the lasers are external triggered, jitter is less than 8~ns, and pulse width less than 10~ns.

\begin{figure}[htb]
\centering
\includegraphics[width=1\textwidth]{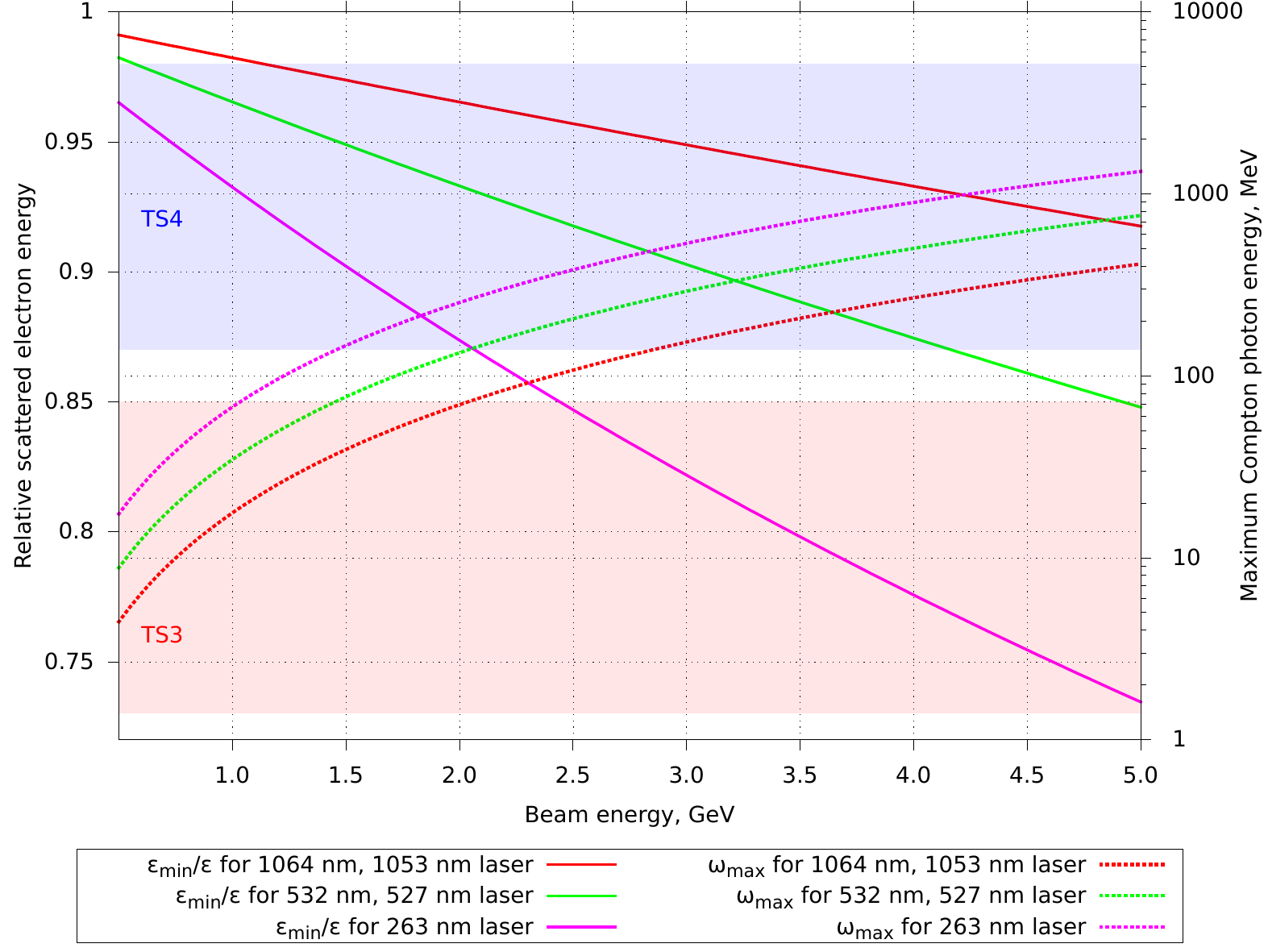}
\caption{Minimum energy of Compton electrons versus the beam energy and TS subsystems energy ranges. Maximum Compton photons energy versus the beam energy}
\label{E_vs_w0}
\end{figure}

These lasers allow to calibrate TS4 subsystem directly by minimal energy of Compton electrons/positrons, see Figure~\ref{E_vs_w0}. The UV-laser (263~nm) would allow to calibrate TS3$^+$ at 2.5~GeV and higher beam energy. Also, the lasers provide Compton $\gamma$-quanta of energies up to 1~GeV which allow to calibrate the BGO calorimeters and measure its energy resolution.

\subsection{Optical System}

At the electron side two laser beams are mixed in one beam with two mirrors. 1064~nm radiation is reflected by a non-transparent mirror to a dichroic mirror transparent for 532~nm light and reflecting for 1064~nm radiation. This beam is focused at the centre of KEDR detector with two positive 30~cm and 100~cm focus length lenses. Before entering the vacuum chamber of VEPP--4M, the beam is reflected by a metallic mirror on a 2-angle movable support with stepping motors. The beam enters the vacuum chamber through 1~cm thickness fused silica window and then is reflected to the electron beam by a metal coated fused silica mirror.

At the positron side 3-wavelength laser emission is attenuated by Glan-Tailor prism in a motorized rotation stage. Then the 3 wavelengths are separated with a Pellin-Broca prism (turns the beam at 90$^{\circ}$) on a horizontal motorized rotation stage. Being reflected by a mirror, one of the wavelengths is focused at the centre of KEDR detector by a motorized laser beam expander. Now we have not motorized laser beam expander for middle-UV radiation, so TS3 can not be calibrated directly. Being reflected by a mirror on a motorized support, the laser beam is inserted into the vacuum chamber through a fused silica window and reflected to the positron beam by a metal coated fused silica mirror.

\begin{figure}[htb]
\centering
\includegraphics[width=0.7\textwidth]{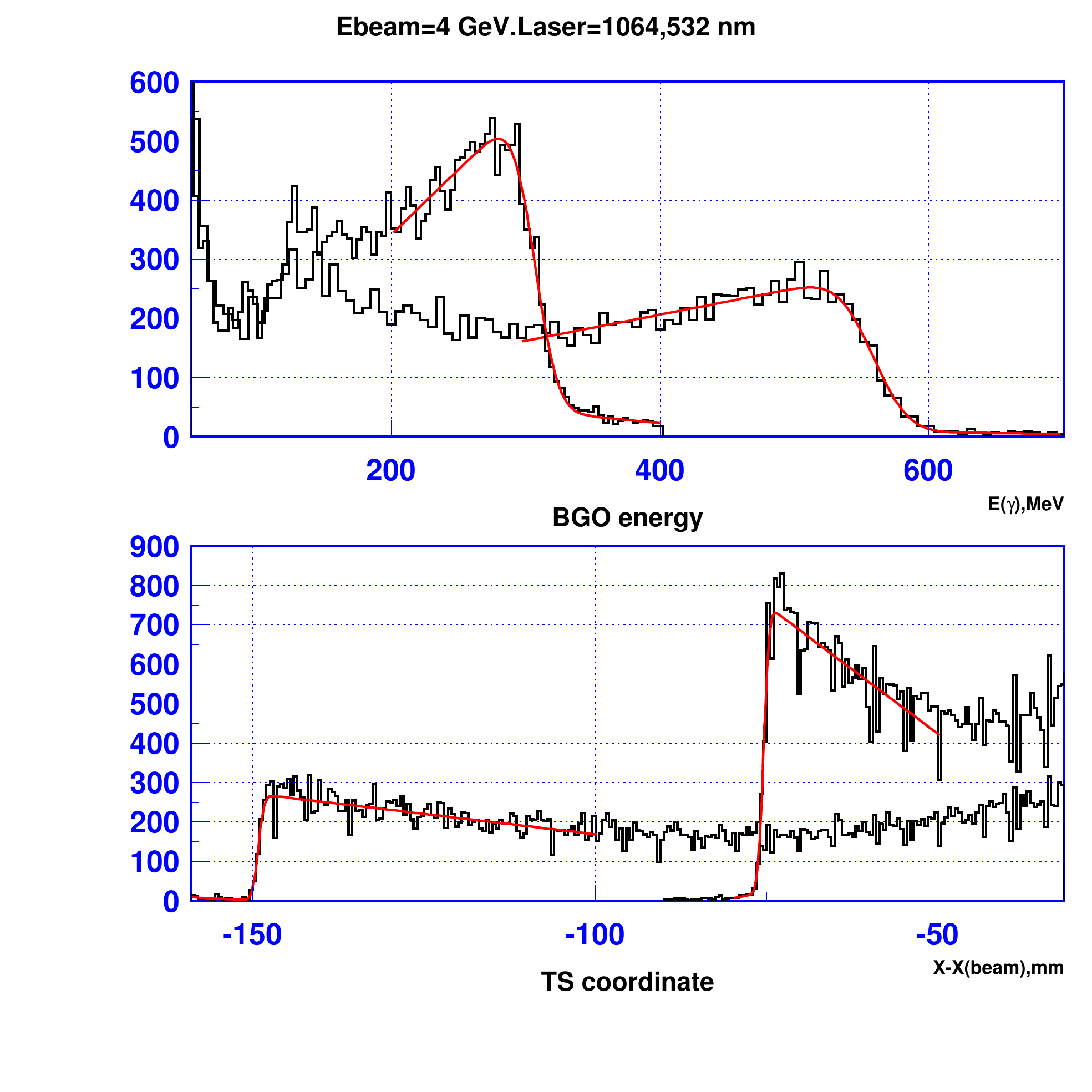}
\caption{Spectra of Compton photons (upper) and Compton electrons (lower) for 2 laser wavelengths}
\label{spectrum}
\end{figure}

\subsection{Trigger and Control}

VEPP--4M beam revolution frequency is 818.924~kHz, meanwhile lasers operate properly with external trigger rates below 50~kHz. The optimal performance of the lasers are gained at frequencies from 1~kHz to 3~kHz. The signal of VEPP--4M RF phase is scaled to 1~kHz, delayed and triggers lasers. This signal also triggers TS and the BGO calorimeter.

Signals from TS4$^{\pm}$ veto counters are used for monitoring Compton events rate.
During VEPP--4M beam run the beam orbit changes. The positions of laser beams are changed by adjusting movable mirrors to provide maximum overlap of the laser beam and the electron or/and positron beam. Compton events rate is used as feedback.

\begin{figure}[htb]
\centering
\includegraphics[width=0.7\textwidth]{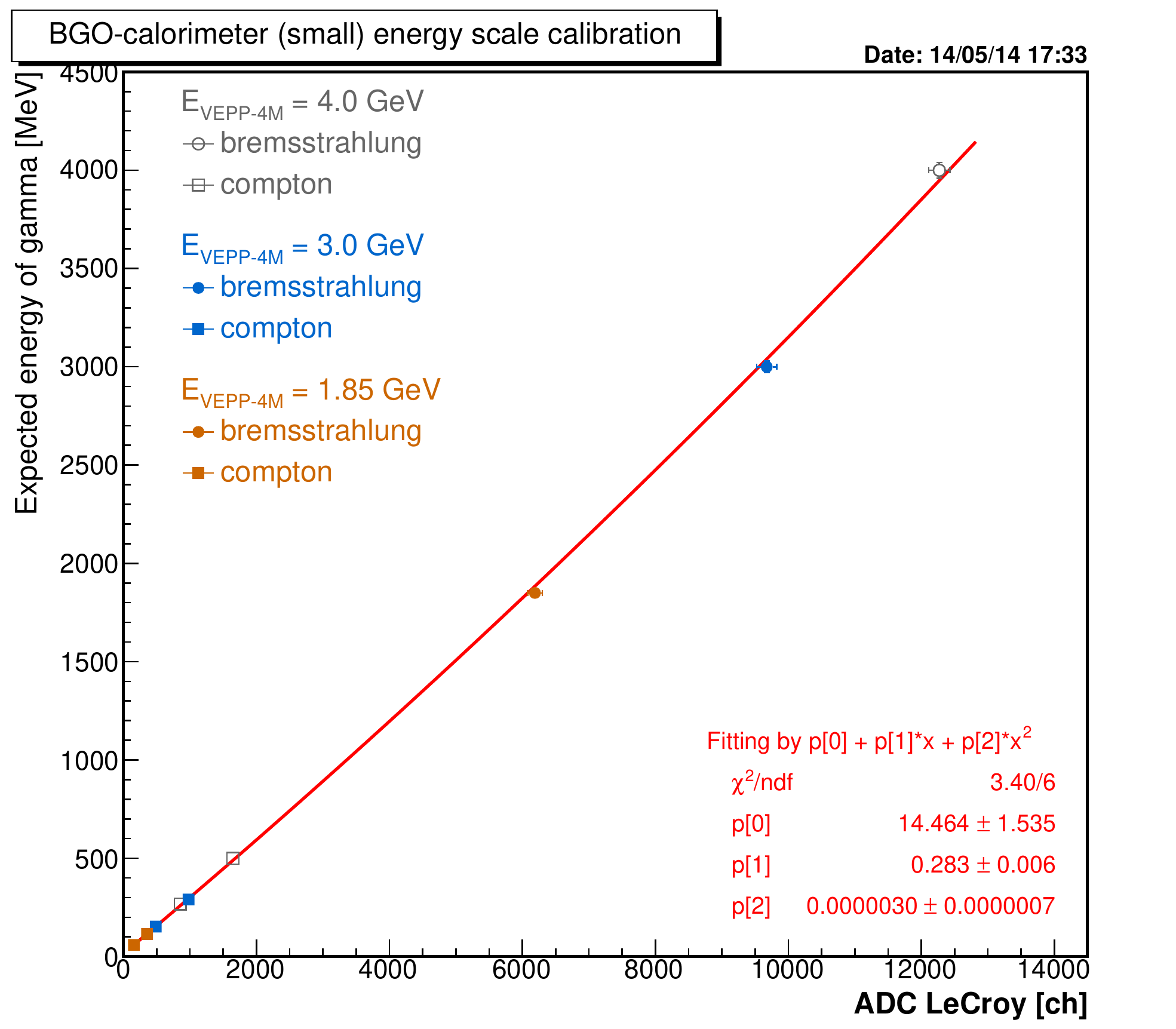}
\includegraphics[width=0.7\textwidth]{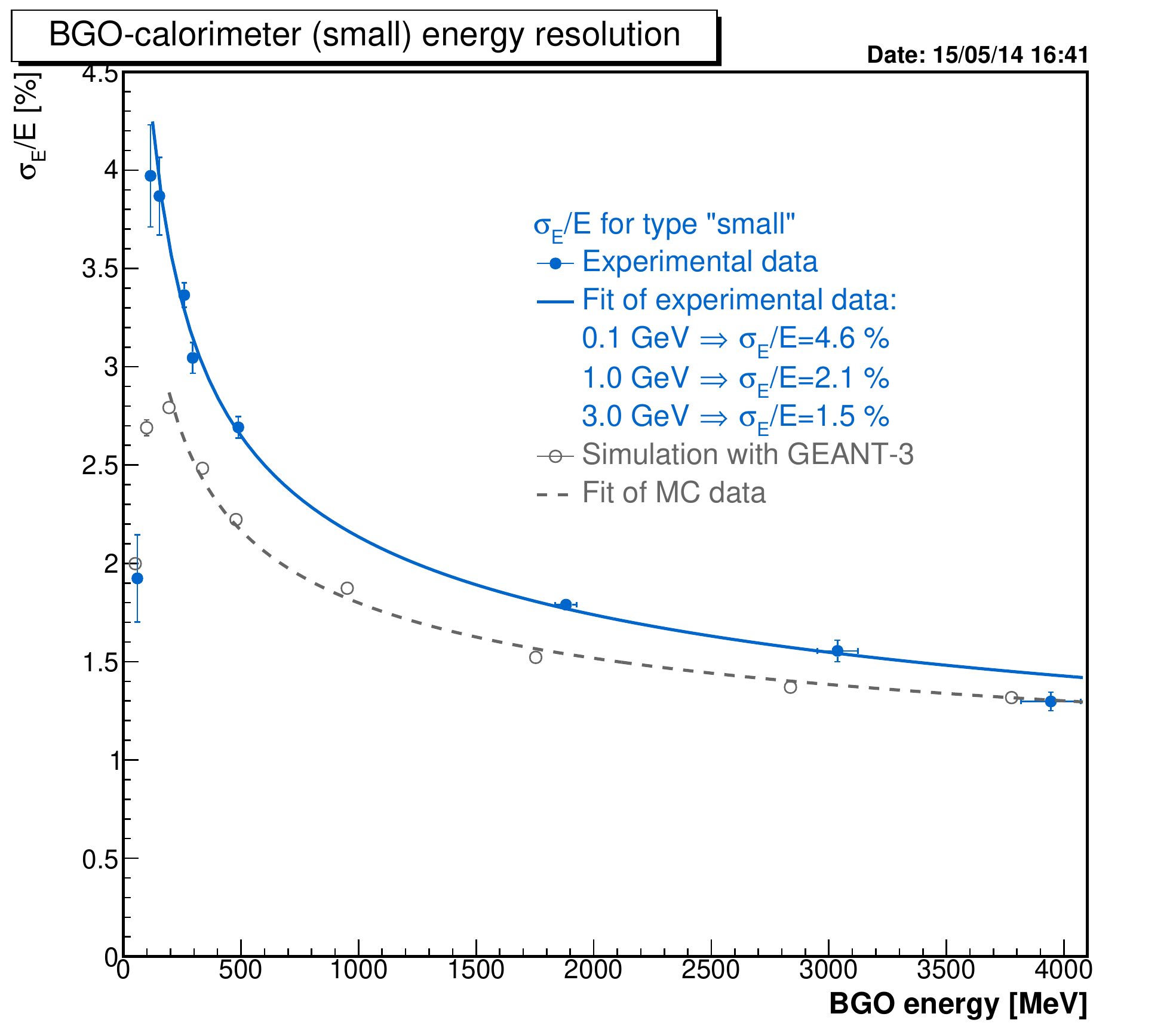}
\caption{Energy scale calibration (left) and relative energy resolution (right) of the BGO calorimeter}
\label{scale_resolution}
\end{figure}

\section{Status and Results}

By the beginning of March 2014 the TS Compton calibration system was ready for routine operation. Both electron and positron laser systems were installed and adjusted. Both Compton electrons (with 1064~nm and 532~nm lasers) and positrons (with 1057~nm and 527~nm lasers) were obtained with high rates.

Preliminary direct calibrations of TS4 and the BGO calorimeter were done. Figure~\ref{spectrum} presents spectra of Compton photons and electrons for 1064~nm and 532~nm lasers, and Figure \ref{scale_resolution} shows the calorimeter energy scale calibration and energy resolution.

\end{document}